\documentclass[11pt]{article}
\usepackage[top = 2 cm, bottom = 2.5 cm, left = 2.5 cm, right = 2 cm]{geometry}
\usepackage{authblk, cite, color, amssymb, amsmath, enumitem}
\usepackage[hypertex, colorlinks = true, linkcolor = blue, citecolor = red]{hyperref}
\usepackage[dvipsnames]{xcolor}

\begin{document}

\title{Towards an information description of space-time}

\author[1,2]{Merab Gogberashvili \thanks{gogber@gmail.com}}
\affil[1]{Javakhishvili State University, 3 Chavchavadze Ave., Tbilisi 0179, Georgia}
\affil[2]{Andronikashvili Institute of Physics, 6 Tamarashvili St., Tbilisi 0177, Georgia}
\maketitle

\begin{abstract}
We attempt to describe geometry in terms of informational quantities for the universe considered as a finite ensemble of correlated quantum particles. As the main dynamical principle, we use the conservation of the sum of all kinds of entropies: thermodynamic, quantum and informational. The fundamental constant of speed is interpreted as the information velocity for the world ensemble and also connected with the gravitational potential of the universe on a particle. The two postulates, which are enough to derive the whole theory of Special Relativity, are re-formulated as the principles of information entropy universality and finiteness of information density.


\end{abstract}


\section{Introduction}

Conventional belief is that the laws of physics are fundamental and can be used as the framework for understanding information processing about the universe. There also exists the opposite view that the laws of physics are just applications of the rules for processing information. The idea that nature is built by information serves as a basis for so-called emergent theories \cite{Car, Ashtekar:1992tm, Connes:1994hv, Jac, Lashkari:2013koa, Ver, Cao:2016mst, Padmanabhan:2009vy, Padmanabhan:2015pza, Padmanabhan:2012gx}. In this approach, even space-time is not an a priori objective reality, but construction of an observer and can be explained in terms of thermodynamic-statistical notions \cite{Amari-1, Amari-2, Caticha}.

The key ingredient in the statistical description of nature is entropy, which captures uncertainties of all kinds and thus allows us to model different aspects of a physical system using a universal mathematical framework. Entropy is a powerful tool in general relativity, thermodynamics, information theories, quantum physics, etc. (see the recent review \cite{Ribeiro-2021}). The concept of entropy is far from being well understood and a more general definition than that introduced by Boltzmann, Shannon, or von Neumann, could exist \cite{Wilde}.

In this paper, we want to model the main features of standard Special Relativity using the informational probabilistic approach \cite{Gogberashvili:2018jkg, Gogberashvili:2021gfh, Gogberashvili:2016llo, Gogberashvili:2013cea, Gogberashvili:2014ora, Gogberashvili:2010em, Gogberashvili:2010ca, Gogberashvili:2008jr, Gogberashvili:2007dw,  Gogberashvili:2016wsa}. Conceptually, it is hard to imagine spacetime as being emerged from some informational pre-geometric variables, we need to generalize the concepts of particles and locality. Locality, one of the assumed properties of classical reality, is incompatible with observation; we already know that quantum theory is nonlocal. In our scenario, the universe is considered as a finite ensemble of non-locally correlated quantum particles, where the kinematics of space-time emerges from the properties of the ensemble. Of course, this is an approximate view, fundamental objects are fields and not particles, we can describe fields using the concept of excitation quanta only in the perturbative regime.

The paper is organized as follows.  After the introduction, in Sec.~\ref{Entropy balance} and \ref{Entropy and action}, we briefly review our earlier results on information formalism and possible links of the entropy to the action. In Sec.~\ref{Information speed} we discuss the possible interpretation of the main fundamental constant of special relativity as the information speed that measures the response time of the world ensemble on the transfer of the information. The Sec.~\ref{Information and metric} is devoted to the information model of geometry. The information (Fisher-Rao) metric is used to calculate the informational difference between measurements. In Sec.~\ref{Relativity postulates} we present arguments that special relativity postulates have natural interpretation in terms of the entropy balance condition of the universe. In Sec.~\ref{Gravity} we speculate if gravity can be described by the entropy conservation condition, based on the fact that thermodynamic entropy and gravity are interconvertible (gravity tries to keep things together and thus tends to lower thermodynamic entropy). Finally, in Sec.~\ref{Conclusions} we summarize our results.


\section{Entropy balance condition} \label{Entropy balance}

Entropy in classical or quantum theories quantifies two seemingly opposite entities: uncertainty (or ignorance) and information (or order). In general, entropy of a physical system contains distinct constituents and there are deep reasons to expect relations between them adopting the principle of total entropy conservation \cite{I-cons},
\begin{equation} \label{S=const}
{\cal S}_{\rm tot} = const~.
\end{equation}
This basic principle of information science constitutes one of the most important elements to our understanding of the quantum universe. According to (\ref{S=const}), after the erasure of some information about a physical system the thermodynamic component of its environment's entropy increases, while the entropy for the whole system, the information coding device plus its environment, is conserved.

When one studies the structure of a physical system and its connections with the whole universe, uncertainty and thus the entropy of the system should decrease. In the limiting case, when all interactions of the system are considered, uncertainties will disappear at all and the total entropy of the system should be zero. So, in our opinion, the total entropy of our finite universe model is not only constant (\ref{S=const}), but can be assumed to be zero, since there are absent observers outside the universe and prior to the big bang as well.

In any measurement process the three systems: an object, memory (or apparatus) and the observer are involved. Then, the zero total entropy of the finite universe can be symbolically written as the sum of information ($I$), thermodynamic (${\cal S}$) and quantum (${\cal H}$) components \cite{Gogberashvili:2016wsa, Gogberashvili:2021gfh}:
\begin{equation} \label{S=0}
{\cal S}_{\rm tot} = I + {\cal S} + {\cal H} = 0~.
\end{equation}
This relation can be understood as the conservation law for all kinds of information in the universe. To fulfill the condition (\ref{S=0}) and at the same time obey the second law of thermodynamics for ${\cal S}$, it seems that in standard definitions of information and quantum entropies we need to include negative components as well \cite{Neg-S-1, Neg-S-2, Cerf:1995sa}.

The expression (\ref{S=0}) seems to be contradictory, since the information and entanglement entropies usually are assumed to be dimensionless. However, the standard definition of thermodynamic entropy, ${\cal S}$, contains Boltzmann's constant $k_B$ and is considered to be a dimensional quantity. We note that in basic physical laws $k_B$ always appears together with the absolute temperature, $T$, and may be considered as its measure in kelvins \cite{Gogber, Atkins, Ka-Ko}. So, it is natural to relate $k_B$ to $T$ and consider $\cal S$ in (\ref{S=0}) as the Shannon type dimensionless information entropy.  This assumption supports the idea that the entropy of statistical mechanics and the information entropy of information theory are basically the same thing \cite{Jay}.

The expression of the total entropy ${\cal S}_{\rm tot}$ as the sum of distinct ingredients, (\ref{S=0}), is most useful in analysis of the physical systems that are separated by horizons. Horizons block information and thus justify the convenience of entropy methods.

The mathematical formulation leading to the association of entropy with any horizon is fairly universal and it does not distinguish between different horizons, like the Rindler horizon in flat space, or a Schwarzschild black hole event horizon, or a de Sitter horizon. It is usually assumed that the event horizon of a black hole can have a purely geometrical definition, while the Rindler horizon is observer dependent. But, similar to what happens in the case of the Rindler frame, an observer plunging into a black hole will have access to different amounts of information (and will attribute different thermodynamic properties to the black hole) compared to an observer who is remaining stationary outside the horizon. In both cases the physical effect of the horizon in blocking information depends on the class of world lines one is considering and it seems necessary to assign an entropy to all horizons.

For physical systems with horizons, different kinds of entropies can be introduced for classically separated space-time regions. According to (\ref{S=0}), evolution of such system (that could be considered as thermodynamically isolated, ${\cal S} = const$) can be described by transitions into each other of information and entanglement entropies (which are observer dependent). For example, for the cases of the Hubble sphere and black holes, the rise of entanglement entropy across their horizons is expected. Then the entropy conservation principle (\ref{S=0}) enables attempts to explain the dark energy of the Universe and to resolve the observed superluminal motion and redshift controversies for black holes \cite{Gogberashvili:2021gfh}.


\section{Entropy and action} \label{Entropy and action}

According to Landauer’s and Brillouin’s principles, the abstract notions of information and entropy can be associated with some physical parameters, like energy or mass \cite{Info, Ilgin:2014dua}. However, there are no unique standards of energy and mass, which appear additive as electric charge. These quantities also have no polarity and usually are taken to be positive. In \cite{Gogberashvili:2016wsa} we had suggested that the convenient physical parameter to measure information (or entropy) is the action $S$ (instead of mass or energy), which like standard entropies usually is considered to be additive and have unique discrete value, $\hbar$. By relating of the thermodynamic entropy with action,
\begin{equation} \label{Statistical-S}
{\cal S}  \sim \frac {S}{\hbar} ~,
\end{equation}
we can develop a concept of energy to account for the amount of change in a system that we observe in time. Any measurement, even a thought experiment, is accompanied by the transfer of at least one quant of the action $\hbar$ and leads to the growth of thermodynamic entropy. Maximal probability has the configuration with highest thermodynamic entropy (i.e. minimal information entropy $I$, as follows from (\ref{S=0})) corresponding to zero interactions, or equilibrium.

If we accept the thesis that action and information are related (\ref{Statistical-S}), the maximal entropy (minimal information) principle \cite{Jay, Pavon:2012qn, B:2017kyi} will become equivalent to the familiar minimal action principle. Physical systems try to evolve so that they do not change the action, that is, the information. However, the principle of least information, that physical systems spontaneously find paths along which they will have as little information emission as possible, is more general, especially for quantum descriptions, since in physical equations one can introduce the information terms (consciousness could also have quantum origins \cite{Ham-Pen}) that correspond to measurement processes and to observers.

The condition of entropy neutrality (\ref{S=0}) also suggests information quantization \cite{Zei}, such as the discreteness of the charge might be thought as the consequence of Gauss' law for the electrically neutral universe -- charge inside of even infinitesimally small volume should be canceled by outside charges. Analogously, there should exist a unit of elementary information -- bit. For a system which entanglement entropy vanishes, the simplified relation (\ref{S=0}) obtains the form:
\begin{equation} \label{S+I=0}
I \left( 1 + \frac {\cal S}{I} \right) = 0~. \qquad \qquad ({\cal H} \to 0)
\end{equation}
The existence of the unit of information suggests that in the process of obtaining information, the related action exhibits discrete features in $\hbar$-s, i.e. upon measurements matter fields behave as particles. So, a bit of information is always associated with a particle and is accompanied by the transfer of elementary action. This assumption, in fact, we already use in standard definition of energy of a particle, $E = \hbar \omega$, the values of $E$ and $\omega$ are different for different particles, while the transfer of the action in any elementary process always remains $\hbar$.


\section{Speed of information} \label{Information speed}

In the world ensemble picture, the universe is considered as the ensemble of entangled quantum particles, where an observer can be considered as a quasi-isolated subsystem. A measurement reduces entanglements of the observed object with the rest of the ensemble. According the generalized Landauer principle \cite{Gogberashvili:2021gfh}, for an observation, in which a bit of entanglement information is lost, into the vacuum the energy $\sim k_B T_{\rm CMB} \ln 2 \sim 1.6 \times 10^{-13}~GeV$ (where $T_{\rm CMB}$ is the present temperature of the vacuum) must be eventually emitted. So, any measurement, even a thought experiment, is accompanied by the transfer of at least one quant of the action $\hbar$ and leads to the change of the configuration of the world ensemble.

Note that, not only there are configurations of the universe which occur more often than other ones, but there are no two configurations with the same weight. The observation of a configuration $k$, changes it and we can assign an ordering (time) in a natural way, $k \sim t$, because any $k'$-configuration contains $k$-configuration if $k' > k$. In this picture, the relation of thermodynamic entropy with information entropies (\ref{S=0}) is the origin of subjective times. So, thermodynamic entropy is related to the time of the event and can serve as the measure of the observer's information, i.e. the entropic arrow of time seems to be an emergent property associated with the transfer of information.

From the introduction of the concept of time for a finite world ensemble follows that there should be a limit to how fast information can move through the universe. Consider a physical system in the volume $V$ that stores some information and the entropy ${\cal S}$. If we neglect entropy supplied by internal sources, the time derivative of the entropy that is contained within this volume should be equal to the flux of the thermal entropy of matter ${\cal S}_m$ through the boundary $A$ \cite{Hartman:2013qma, Liu:2013iza, Liu:2013qca},
\begin{equation} \label{dot S}
\frac {d{\cal S}}{dt} = {\cal S}_m v\,\frac {A}{V} ~.
\end{equation}
In this equation it appears the quantity with the dimension of velocity $v$ (information speed), which controls the rate of information growth in thermalizing states. From (\ref{dot S}) follows that, in a simple model, the entropy of the system, initially contained within the volume $V$ and entangled with outside matter, grows linearly in time (entanglement tsunami) \cite{Kuwahara:2019rlw},
\begin{equation} \label{S-th}
{\cal S} = {\cal S}_{m} v t \,\frac {A}{V} ~.
\end{equation}

To estimate the information speed $v$ for the entire universe let us write the equation (\ref{dot S}) for the Hubble horizon $R_H = c/H$. Using the black hole type entropy for the Hubble sphere \cite{Bekenstein:1973ur},
\begin{equation} \label{Bekenstein}
{\cal S}_H = \frac {A_H c^3}{4G\hbar} = \frac{\pi c^5}{G \hbar H^2} ~,
\end{equation}
were $G$ is the Newton constant, and the Euler relation,
\begin{equation}\label{Gibbs-Duhem}
\rho + \frac {p}{c^2} = k_BT_H \,\frac {{\cal S}_m}{V_H} ~,
\end{equation}
where $\rho$ stand for matter density, $p$ represents pressure density and
\begin{equation}
T_H = \frac {\hbar H}{2\pi k_B c^2}
\end{equation}
denotes the de Sitter temperature associated to the horizon \cite{Hawking:1975vcx}, the relation (\ref{dot S}) leads to the standard Raychaudhuri, or the second Friedmann equation \cite{Gogberashvili:2016wsa, Gogberashvili:2021gfh},
\begin{equation} \label{dotH}
\frac {dH}{dt} = - 4\pi G \left(\rho + \frac {p}{c^2}\right) ~,
\end{equation}
but only if
\begin{equation} \label{v=c}
v = c~.
\end{equation}
So, in statistical formalism, the speed of light $c$ can be understood as the world ensemble parameter that corresponds to the information velocity. This information speed measures the response time of the world ensemble on the transfer of one bit of information, i.e. exchange of one particle.


\section{Information and metric} \label{Information and metric}

In an information probabilistic approach space is not an a priori objective reality and the metric represents the potential knowledge (or ignorance) of an observer. Geometry does not exist without matter, in practice we measure excitations of quantum fields whose actual locations have no meaning. An observer associates to each particle the entropy ${\cal S}(x^i)$ and the probability distribution,
\begin{equation}
P(x^i) \sim e^{-{\cal S}(x^i)}~,
\end{equation}
labeled by some set of informational coordinates $x^i$.

To find the dimension of the information space, or the number of independent coordinates $x^i$, consider the simplified model of the world ensemble consisting of $N$ identical particles (generalization for many spices is obvious). The total action of the system, $S_{\rm U}$, which we use in the definition of the total thermodynamic entropy (\ref{Statistical-S}), contains the factor $N^3$ \cite{Gogberashvili:2013cea, Gogberashvili:2014ora, Gogberashvili:2010em},
\begin{equation} \label{A-U}
S_{\rm U} \sim N^3 \hbar ~,
\end{equation}
and not the factor $\sim N$, as in a local model. It is known that the maximal entropy of a random variable with $n$ realizations is $\ln n$. The maximal entropy of the model universe appears to be
\begin{equation}
{\cal S}_{\rm U}  \sim \ln N^3 = 3 \ln N~.
\end{equation}
Then, in local measurements an observer obtains three copies of the entropy, $\sim \ln N$, which is used to define information distances. So, in the definition of a particle distribution function, $P(x^i)$, the number of the information coordinates, $x^i$, three may be required ($i=1,2,3$).

For a macroscopical physical system of mass $M$ and the size $R$ we have two Bekenstein’s entropy bounds: the universal bound (independent of the existence of gravity), ${\cal S}_{\rm max} \sim MR$ \cite{Bekenstein:1980jp}, and the area-law entropy bound (\ref{Bekenstein}) (with gravity), ${\cal S}_{\rm max} \sim R^2/G$ \cite{Bekenstein:1973ur}. For theories that saturate unitarity, these bounds are equal \cite{Dvali:2020wqi, Dvali:2019jjw, Dvali:2019ulr}, and lead to the relation
\begin{equation} \label{2GM/R}
\frac {2GM}{R} = c^2~,
\end{equation}
which serves as the definition of Schwarzschild radius and also relates with each other the fundamental parameters $c$ and $G$. Note that (\ref{2GM/R}) in addition is equivalent to the critical density condition in relativistic cosmology, $\rho_{c} = 3H^2/8\pi G$.

According to (\ref{Bekenstein}), the maximal entropy of a region of the size $R$, independently of its matter content, is ${\cal S}_{\rm max} \sim R^2$. Thus, an observer naturally associates ${\cal S}$ with the notion of a spherical volume of the radius,
\begin{equation} \label{R2=S}
R^2 = g_{ij} x^ix^j \sim {\cal S}(x^i)~, \qquad (i,j = 1,2,3)
\end{equation}
where the probability distribution $P(x^i)$ becomes uniform, or is the least informative.

In continuous measurements it is difficult to distinguish quantum particles, the information distance between two neighboring probability distributions, $p(x^i)$ and $p(x^i + dx^i)$, is given by the variance \cite{Amari-1, Amari-2, Caticha},
\begin{equation} \label{Inf-metric}
dl^2 = g_{ij} dx^idx^j ~. \qquad (i, j = 1,2,3)
\end{equation}
Consider the entropy ${\cal S}(x^i, x'^i)$ of one distribution $p(x^i)$ relative to another distribution $p(x'^i)$, which attains an absolute maximum at $x^i = x'^i$. In terms of ${\cal S}(x^i, x'^i)$ the information metric $g_{ij}$ in (\ref{Inf-metric}) can be defined as
\begin{equation} \label{g=S}
g_{nm} = - \frac {\partial {\cal S}(x^i, x'^i)}{\partial x'^n \partial x'^m}~, \qquad (i, n, m = 1,2,3)
\end{equation}
so that \cite{Amari-1, Amari-2, Caticha}
\begin{equation}
{\cal S} (x^i + dx^i, x^i) = - \frac 12 dl^2~.
\end{equation}
The function ${\cal S}(x^i)$ contains all information about a space since it allows one to recover its metric. As a consequence, the notion of distance, including the metric as its infinitesimal version, can be replaced by the notion of observability, smaller is the relative entropy of distribution for two particles shorter is distance between them. A small value of (\ref{Inf-metric}) means that particles at the points $x^i$ and $x^i + dx^i$ are difficult to distinguish. We can also invert the logic and assert that the two points $x^i$ and $x^i + dx^i$ must be very close together because they are difficult to distinguish.

The metric (\ref{Inf-metric}) is known as the Fisher-Rao metric, or the information metric, which can be used to calculate the informational difference between measurements. The informational coordinates $x^i$ are arbitrary, one can relabel distributions of particles. It is then easy to check that $g_{ij}$ are the components of a tensor and that the distance $dl^2$ is an invariant, a scalar under coordinate transformations. It is important that the metric tensor $g_{ij}$ on the manifold of probability distributions is unique: there is only one metric (up to rescaling) that is invariant under sufficient statistics, i.e. another observer with the same thermodynamic entropy has no additional information \cite{Amari-1, Amari-2, Caticha}.

The information distance (\ref{Inf-metric}) is dimensionless, $g_{ij}$ measures distinguishability in units of the local uncertainty implied by the distribution $p(x^i)$. So, information geometry allows us to describe the conformal geometry of space - the local shapes but not absolute local sizes. The scale of distance turns out to be a property of the models we employ to describe it. One possible choice of gauge would be to choose the unit of length so that the evolving three-dimensional information manifold of distributions generates a four-dimensional space-time. The conditions for such a space-time gauge have been proposed, for example, in the context of Machian relational dynamics \cite{Gomes:2010fh}.


\section{Relativity postulates} \label{Relativity postulates}

In this paper we consider the universe as a unite system and want to describe emergence of its local relativity properties in terms of information quantities. The Riemannian space-like metrics of information geometry (\ref{Inf-metric}) do not reproduce the light-cone structure of space-time; some additional ingredient is needed. According to (\ref{S=0}), changes in information and entanglement entropies for a thermodynamically isolated physical system (coordinate transformations), leads to deformations in thermodynamic entropy associated with time evolution. This relation of thermodynamic and information entropies is analogous to the relativistic transformations. Quantum theory does not treat space and time on the same footing. Using the dimensional parameter of information speed (\ref{v=c}), the information space (\ref{Inf-metric}) can be considered as a three-dimensional spacelike 'surface' embedded in four-dimensional space-time and thus $c$ introduces the scale for the information distance (\ref{Inf-metric}). Having decided on a measure of information distance, $dl$, we can now also measure angles, areas, volumes and all sorts of other geometrical quantities. The introduction of notions of space-time and the relation of thermodynamic entropy to action integral, moves information theory from abstract mathematics to physics.

Now we are ready to try to translate relativity theory in information terms. In the Special Theory of Relativity the law of inertia has no known origin \cite{Feynman}. This theory also says nothing about why an inertial system exists, which is required to describe relative motions and inertial observers. In the world ensemble picture, all entangled particles are employed in the definition of the fundamental frame. Also, homogeneity and isotropy of the universe (with a huge number of particles) is obvious for a small system. In spite of the introduction of the preferred cosmological frame, the world ensemble model can imitate basic features of relativity theory locally, such as the relativity principle, the local Lorentz invariance and the so-called signal locality (matter cannot propagate faster than $c$).

In our model relativity emerges due to the existence of the class of 'inertial' observers with the same informational, but different thermodynamic and entanglement entropies. From the assumption (\ref{S=0}) it follows that for observers with the constant information entropy about the universe, $I = const$, the changes in thermodynamic and entanglement entropies should cancel out,
\begin{equation} \label{dS=-dH}
  d{\cal S} = -d{\cal H} ~.
\end{equation}
When in the finite universe we consider a physical system with the marginal quantum entropy ${\cal H}_1$, we divide the universal ensemble in two parts. If the entropy of an overall state is zero, ${\cal H}_U = 0$, the marginal entropies of a pure multi-partite state are equal \cite{Wilde},
\begin{equation} \label{S1=S2}
{\cal H}_1 = {\cal H}_2~,
\end{equation}
where ${\cal H}_2$ is the marginal quantum entropy of the rest of the universe. So, the entanglement entropy of a system ${\cal H}_1$ should equal to the accompanied entropy of the universe ${\cal H}_2$. The relation (\ref{S1=S2}) is the special case of the Araki-Lieb inequality \cite{Araki-Lieb}:
\begin{equation} \label{S1+S2}
|{\cal H}_1 - {\cal H}_2| \leq {\cal H}_U \leq {\cal H}_1 + {\cal H}_2~.
\end{equation}
If the composite system is in a pure state, ${\cal H}_U = 0$, the left-hand side of the inequality (\ref{S1+S2}) reduces to (\ref{S1=S2}). The property (\ref{S1+S2}), which according to (\ref{dS=-dH}) can also be translated for the thermodynamic entropies $\cal S$, is the radical departure from the classical world, where the joint entropy ${\cal H}_U$ is never less than one of the marginal entropies.

Equality in the right-hand side of (\ref{S1+S2}) holds when both systems are independent and the total entropy is just a sum of the sub-entropies. Physically, this implies that the entropy of the composite system ${\cal H}_U$ is maximized when its components are uncorrelated; we have more information (less uncertainty) in an entangled state than if the two states are treated separately. So, some information that is important to describe a physical system is stored by the entangled states of the entire world ensemble and cannot be obtained from the local Lagrangians. This is the manifestation that the information-probabilistic approach is more general than standard variational principles.

For two observers with zero overall entropies and the same information entropy, $I_1 = I_2$, the sums of thermodynamic and entanglement entropies are equal. For the case of constant but different thermodynamic and entanglement ingredients we have an 'inertial' observer with the same information speed (\ref{v=c}) and the same information entropy about the universe. Thus, for any physical system that can be treated as isolated, there exists a special state (inertial frame) for which the rest of the universe looks similar. Thus, the local relativistic invariance is a fictitious symmetry that has been artificially imposed in physical models to account the observer dependence of geometry. This fictitious symmetry disappears for large, cosmological scale systems, i.e. the relativity principle is subjective and local. Entropy approach leads to a new dimension of inertial frames of reference and the transformation between them. It also removes the concepts ‘rest frame’ and ‘moving frame’ and emphasizes the importance of observer and observation.

Let us demonstrate subjective properties of the relativity principle using our model universe of $N$ quantum particles in the equilibrium. Consider an elementary fluctuation with action that is written as the product of two quantities,
\begin{equation} \label{A=Et}
S = Et~,
\end{equation}
one of which, the energy $E$, does not vary during transition from equilibrium to excited state and thus can be attributed to the fluctuation itself. The quantity $E$ is subjective and is minimal from an observer’s point of view (in the observer’s frame). The quantity $t$ (time) describes the change of fluctuation (when the model universe bounces back to equilibrium) and can be used as the kinematical parameter describing the evolution of the ensemble. In the linear case (\ref{S-th}), for a multi-particle fluctuation, an observer can still write corresponding action in the form (\ref{A=Et}), introducing some fictitious energy $E'$ and the same time parameter $t$,
\begin{equation} \label{A+extra}
S' = Et + S_{\rm extra} = E't~.
\end{equation}
The extra term, $S_{\rm extra}$, alters total thermodynamic entropy of the observer and leads to the variance of the distribution functions of particles of the ensemble in the expression of the emerged distance (\ref{Inf-metric}). Thus extra term in (\ref{A+extra}) can be interpreted as describing motions,
\begin{equation} \label{}
S_{\rm extra} = p_ix^i~,
\end{equation}
i.e. the relation (\ref{A+extra}) represents the relativity principle in 'space-time'.

Now let us try to re-formulate the Special Relativity axioms in information terms (see also \cite{Gogberashvili:2022wsh, DelSanto:2021aem}). Consider the following two axioms:
\begin{enumerate}[label= (\Alph*)]
\item Principle of informational entropy universality: {\it The laws of physics have the same form for the observers with the same information, but different thermodynamic and entanglement entropies;}
\item Principle of finiteness of information density (Bekenstein bound \cite{B-bound}): {\it A finite volume of space can only contain a finite amount of information.}
\end{enumerate}
From (B) (that replaces the requirement of invariance of $c$) one infers that only a finite amount of information can be transmitted in a finite time, otherwise this would require to 'move' an infinite volume of space. Hence, the speed of propagation of information (\ref{v=c}) necessarily needs to be finite too. In the finite universe there should exist maximal speed, corresponding to the transfer of all information of the universe in its lifetime. Then, from (A) (analog of the principle or relativity), this velocity must be the same for all 'inertial' observers. The former considerations are enough to derive the whole theory of Special Relativity, with the additional assumptions of homogeneity of space and time and isotropy of the world ensemble.

According to (\ref{S-th}), thermodynamic entropy of a moving particle increases as,
\begin{equation} \label{S=vt}
d{\cal S} \sim d(v t) ~.
\end{equation}
Then, for a system which entanglement entropy vanishes,
\begin{equation}
dI \approx - d{\cal S} ~.  \qquad \qquad ({\cal H} \to 0)
\end{equation}
So, Lorentz contractions of the induced space-time coordinates appear to be the direct consequence of the relation (\ref{S+I=0}). It is known that, if we define inertial observers with classical notions of space-time coordinates, the standard Lorentz transformations can be derived using only the relativity principle, supplemented by the assumptions of homogeneity, isotropy and smoothness \cite{Lorentz-1, Lorentz-2}. The existence of a universal speed also follows as a consequence of the relativity principle.

Another physical consequence of the principle of finiteness of information density (B) is that, in general, physical quantities (say the position of a particle at a certain moment) do not have perfectly determined values at every time (or, alternatively, that the truth value of certain empirical statements, such as a particle location at a certain instant, is indeterminate). Hence, upholding the principle of finiteness of information density gives us a hint that physics should be at the same time indeterministic and relativistic. These two views are compatible if one regards (in)determinacy itself as relative.


\section{Gravity} \label{Gravity}

In the last section we want to discuss some properties of our finite world ensemble model and the zero-entropy assumption (\ref{S=0}) that can help in the construction of future information theory of gravity.

Because of allowing non-local correlations and negative components in the expressions of information and entanglement entropies, some particles of the world ensemble may stumble upon one another and get ordered, i.e. due to increase in $I$ and ${\cal H}$, according to (\ref{S=0}), the thermodynamic entropy ${\cal S}$ may decrease at that particular portion of space. This property we can connect with gravity, it tries to keep things together through attraction and thus tends to lower thermodynamic entropy \cite{Grav.Ent-1, Grav.Ent-2, Grav.Ent-3, Grav.Ent-4}. Gravity (encoded in $I$ and ${\cal H}$ components) and thermodynamic entropy have a reciprocal existence, meaning that, according to (\ref{S=0}), increase in one causes decrease in the other. Thus, thermodynamic entropy and gravity are inter-convertible. But quantum measurements and gain of information are irreversible processes and the conversion from gravity into thermodynamic entropy seems to be unidirectional, which we call time’s arrow.

It is known that the Einstein equations can be derived using the ideas in holography and of the entropy area relation (\ref{Bekenstein}) \cite{Jac, Lashkari:2013koa}. By combining general thermodynamic considerations with the equivalence principle, equations for gravity also can be written as a single scalar relation, which has the natural interpretation as the balance of gravitational and matter heat densities, in the spirit of the first law of thermodynamics \cite{Padmanabhan:2009vy}. To describe gravity in terms of information quantities within our model we note that the total entropy of the quantum ensemble of $N$ identical particles can be written as
\begin{equation} \label{sum S}
{\cal S}_{\rm tot} = {\cal S} + \sum {\cal S}_N~,
\end{equation}
where ${\cal S}_N$ are thermodynamic entropies of the members of the ensemble. This relation implies that, if after the subtraction of $\sum {\cal S}_N$, from the total entropy of the universe ${\cal S}_{\rm tot}$, we are left with non-zero entropy $\cal S$, then gravity is present. If we assume that an observer is in thermal equilibrium with the rest of the universe, his thermodynamic entropy $\sum {\cal S}$ tends to maximum and, according to (\ref{sum S}), the gravity (or acceleration) has to vanish for him. So, maximal entropy (or minimal information) triggers the gravity to disappear.

Another plus of the world ensemble model is that, according to (\ref{2GM/R}), the ensemble parameter $c$ can be understand also as the non-local gravitational potential of the entire universe of the mass $M_U$ acting on each particle:
\begin{equation} \label{Phi}
c^2 = - \Phi = \frac {2GM_U}{R_H}~.
\end{equation}
The 'universal' gravitational potential $\Phi$, and thus $c$, can be regarded as constants, since for a small system the world ensemble is isotropic and homogeneous. This allows us to relate the origin of inertia (or rest energy) of a particle from the world ensemble to its gravitational interactions with the whole universe:
\begin{equation} \label{mPhi}
E = mc^2 = - m\Phi~.
\end{equation}
Thus, in the world ensemble scenario, the entire ensemble is engaged in any local interaction, what represents some kind of Mach's Principle \cite{Gogberashvili:2010em, Gogberashvili:2010ca, Gogberashvili:2008jr, Gogberashvili:2007dw}.

Note also that the assumption that the whole universe is involved in local interactions, can effectively weaken the observed strength scale of gravity by a factor related to the number of all particles and might resolve the hierarchy problem in particle physics \cite{Gogberashvili:2008jr}.

Consider a system that contains a sufficiently large number of particles $n \le N$, i.e. has its own large thermodynamic entropy ${\cal S}$. For such object the relation (\ref{S+I=0}) should contain the extra term
\begin{equation}
\frac {\cal S}{I} \sim \frac {n}{N}~.
\end{equation}
This locally reduces the value of the 'universal' gravitational potential (\ref{Phi}),
\begin{equation} \label{c'}
c'^2 \sim \frac nN c^2 \sim (1-f) c^2~,
\end{equation}
which is equivalent to the appearance of the local gravitational potential,
\begin{equation}
f \sim \frac {N-n}{N}~.
\end{equation}

Let us also try to explain the equivalence principle for the world ensemble model. For a small physical system, in the simple linear time dependent case (\ref{dot S}), the formula for entropy grows contains the parameter with dimension of velocity, $v = c$. When entropy of the system is a more general function of time, we can still use the expression (\ref{dot S}), but with time-dependent information velocity $v (t)$. This leads to the anisotropic deformation of the fundamental speed parameter, as in (\ref{c'}), which imitates gravitation interaction locally and is a representation of the equivalence principle.

At the end let us note that the idea behind our model is not to build up the spacetime from some pre-geometric variables, like ‘atoms of spacetime’ \cite{Padmanabhan:2015pza}, but to consider the universe as a unite system (in the spirit of the General Relativity) in terms of information quantities. In this approach we have no problems with explanation of emergence of space around local gravitating objects (direct experience tells us that space around them is pre-existing) \cite{Padmanabhan:2012gx}, since any finite body can be considered as the subsystem entangled with the universe.

Generic problem of emergent theories is also the treatment of time differently from space, that runs counter to the spirit of relativistic invariance and general covariance \cite{Padmanabhan:2012gx}. This difficulty disappears in the cosmological context, since the existence of a cosmological preferred frame justifies considering time differently from space. In our model the relativistic invariance emerges locally by considering equivalent classes of observers with the same information entropy about the rest of the universe (see the Sec.~\ref{Relativity postulates}). At the same time, for any observer the information about the universe can be accounted for by the second, cosmological metric with the global ($1+3$)-splitting. So, we have some kind of biometric model, one metric describes information of the observers about the universe and the other determines the geometry we probe locally. This kind of model not only gain the advantage of ordinary bimetric models that the connection (defined as the difference of two connections with respect of both metrics) is a tensor and useful to define conserved energy-momentum tensor for gravity (for bimetric models see, for example, the reviews \cite{Schmidt-May:2015vnx, Clifton:2011jh}), but seems to be free of ghost in matter sector, since the information metric is non-dynamical and couples only to the observer.


\section{Conclusions} \label{Conclusions}

To conclude, in this paper we attempted to describe emerged geometry in terms of informational quantities. Such an approach is more general, since in physical equations one can introduce the information terms that correspond to measurement processes and to observers.

We consider the universe as a finite ensemble of non-locally correlated quantum particles. As the main dynamical principle, the assumption of conservation of the sum of all kinds of entropies (thermodynamic, quantum and informational) was used. Instead of mass or energy, we relate thermodynamic entropy to action that moves information theory from abstract mathematics to physics. The fundamental constant of speed is interpreted as the information velocity for the finite world ensemble of all quantum particles. This constant also represents the gravitational potential of the universe acting on each particle of the ensemble. The two postulates, which are enough to derive the whole theory of Special Relativity, are re-formulated as the principles of information entropy universality and finiteness of information density.


\section*{Acknowledgments:}

This work was supported by Shota Rustaveli National Science Foundation of Georgia (SRNSFG) through the grant DI-18-335.


\section*{Data Availability:}

No new data were created or analyzed in this study. Data sharing is not applicable to this article.


\end{document}